\documentclass[twocolumn,aps,showpacs,showkeys,superscriptaddress,notitlepage,longbibliography]{revtex4}
\usepackage[colorlinks=true,urlcolor=blue,citecolor=blue,linkcolor=blue]{hyperref}
\usepackage{graphicx}  % needed for figures
\usepackage{dcolumn}   % needed for some tables
\usepackage{bm}        % for math
\usepackage{amssymb}
\usepackage{amsfonts}
\usepackage{doi}
\usepackage{verbatim}
\usepackage{epstopdf}

\usepackage{amsmath}
\usepackage{braket}
\usepackage{hyperref}
\DeclareMathOperator{\Tr}{Tr}

\begin{document}

\title{ Emergent $Z_2$ Topological Invariant and Robust Helical Edge States in Two-Dimensional Topological Metals}
\author{Chui-Zhen Chen}
\affiliation{Institute for Advanced Study and School of Physical Science and Technology, Soochow University, Suzhou 215006, China.}
\author{Hua Jiang}
\affiliation{Institute for Advanced Study and School of Physical Science and Technology, Soochow University, Suzhou 215006, China.}
\author{Dong-Hui Xu}
\affiliation{Department of Physics, Hubei University, Wuhan 430062, China. }

\author{X. C. Xie}\thanks{xcxie@pku.edu.cn}
\affiliation{International Center for Quantum Materials, School of Physics, Peking University, Beijing 100871, China}
\affiliation{CAS Center for Excellence in Topological Quantum Computation,
University of Chinese Academy of Sciences, Beijing 100190, China}
\affiliation{Beijing Academy of Quantum Information Sciences, West Bld.3,
No.10 Xibeiwang East Rd., Haidian District, Beijing 100193, China}
\begin{abstract}
In this work, we study the effects of disorder on topological metals that support a pair of helical edge modes deeply embedded inside the gapless bulk states. Strikingly, we predict that a quantum spin Hall (QSH) phase can be obtained from such topological metals without opening a global band gap. To be specific, disorder can lead to a pair of robust helical edge states which is protected by an emergent $Z_2$ topological invariant, giving rise to a quantized conductance plateau in transport measurements. These results are instructive for solving puzzles in various transport experiments on QSH materials that are intrinsically metallic. This work also will inspire experimental realization of the QSH effect in disordered topological metals.

\end{abstract}

\maketitle
%\setpagewiselinenumbers

\section{Introduction}
A quantum spin Hall (QSH) insulator, or a two-dimensional topological insulator, is a symmetry-protected topological phase of matter that is insulating in its interior but supports gapless helical states confined to its edges~\cite{KaneRMP2010,ZhangRMP2011,KanePRL2005,KanePRL05,Andrei2006}. Due to spin-momentum locking in helical edge states, elastic backscattering is forbidden without breaking time-reversal symmetry ~\cite{KanePRL2005,KanePRL05,Andrei2006,Konig2007,Roth2009}. This gives rise to ballistic electron transport and consequently a hallmark quantized edge conductance of 2$e^2/h$ in transport measurements.
However, truly quantized edge conductance is difficult to observe.
 The key challenge is that the bulk states in many QSH candidates are actually metallic~\cite{Du2011,Qian2014,Song2018,Wang2016}, which spoils the quantization of the edge conductance.
Indeed, the helical edge modes were only identified in HgTe/CdTe quantum wells (QWs)~\cite{Konig2007,Roth2009} and InAs/GaSb QWs~\cite{Liu2008,Du2011,Du2014}, as well as recently in the 1T$^\prime$-WTe$_2$ monolayer~\cite{Qian2014,Fei2017,Tang2017,Wu2018,Song2018}.
Nevertheless, the origin of the quantized edge conductance in InAs/GaSb QWs~\cite{Liu2008,Du2011,Du2014} and 1T$^\prime$-WTe$_2$ monolayer remains controversial \cite{Qian2014,Fei2017,Tang2017,Wu2018,Song2018}.
Transport measurements show sizable bulk conduction in the intrinsic InAs/GaSb QWs ~\cite{Du2011} while scanning tunneling microscope (STM) studies reveal a metallic bulk band structure in the 1T$^\prime$-WTe$_2$ monolayer \cite{Song2018}.

Disorder plays an essential role in electron transport phenomena in one or two dimensions due to Anderson localization \cite{Anderson1958,Anderson1979}. Nevertheless, the helical edge states in QSH insulators are not subject to Anderson localization for weak  non-magnetic disorder. That is because the $Z_2$ topological invariant, distinguishing a QSH insulator from a normal insulator, does not change until the bulk gap closes \cite{KanePRL2005,KanePRL05,Onoda2007,Shen2009,Jiang2009,Groth2009,Juntao2012,Dongwei2012,GA2013,Yamakage2013,Yan-Yang2014}.
%Currently, the studies of disorder effect on QSH effect are mostly concentrated on the system with a intrinsic band gap
Unfortunately, in many QSH materials the valence band overlaps the conductance band as it does in metals, and the helical edge states are embedded inside the bulk states, which is dubbed topological metals. Apparently, the robustness of a gapped topological system against disorder is not directly applicable, since there is no band gap to sustain the $Z_2$ invariant. One would naively expect that the edge and bulk states are mixed and localized indistinguishably in the presence of disorder, and there is no way to realize the QSH effect with a quantized edge conductance in these materials. However, in the following we demonstrate that this is not the case for topological metals.
%As a result, it is crucial to study disorder effect of the topological metals, which possess helical edge modes embedded inside gapless bulk states.

%This raise an important question on the fate of helical edge states in the gapless topological system in presence of nonmagnetic %disorder.

\begin{figure}[bht]
\centering
\includegraphics[width=3.4in]{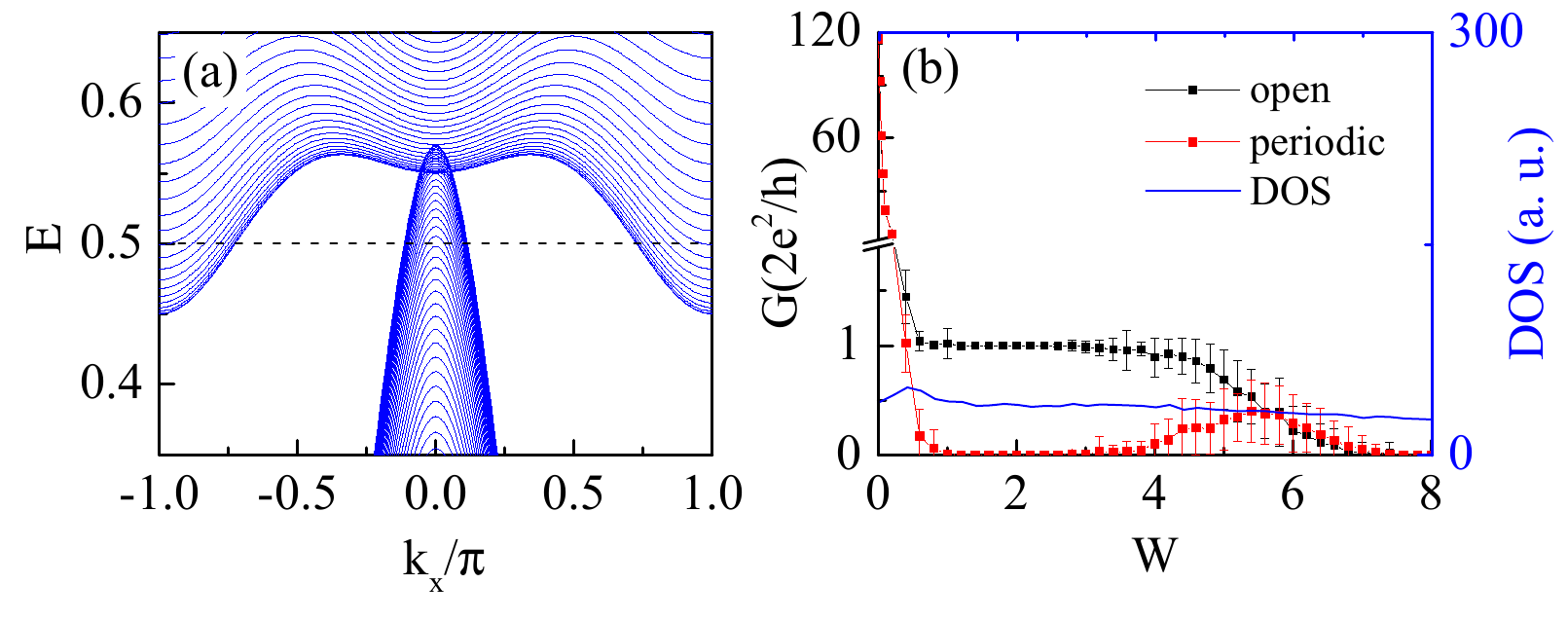}
\caption{(Color online). (a) The energy band structure of a nanoribbon with the width $L=256$ for the modified BHZ model. The dashed line indicates the Fermi energy at $E=0.5$. (b) Two terminal conductance $G$ versus the disorder strength $W$ at the Fermi energy $E=0.5$ with a periodic (open) boundary condition in the $y$ direction. The conductance is averaged over $48$ disorder configurations for the sample size $512\times 512$. The conductance fluctuations are shown as error bars. For comparison, DOS in an arbitrary unit (a.u.) is shown in the same panel.
\label{fig1} }
\end{figure}

In this work, we study the disorder effect on topological metals which are described by a modified Bernevig-Hughes-Zhang (BHZ) model.
In the clean limit, the system is a topological metal in which the valence and conductance bands overlap with each other and there is no band gap [see Fig.~\ref{fig1}(a)]. Strikingly, in the presence of disorder, a QSH phase characterized by the $Z_2$ topological invariant emerges by opening up a mobility gap. The emergent QSH phase is featured by a quantized two-terminal conductance and localized bulk states with a finite density of states (DOS) [see Fig.~\ref{fig1}(b)]. These results provide an alternative way of understanding  the current experiments on QSH materials and will also stimulate further studies to realize QSH effect in disordered topological metals.

%\C{ We demonstrate that the edge states in the disordered topological metals are topologically protected by the spin Chern number by combining the finite-size scaling of the two-terminal conductance, spin-polarized Hall conductance and localization length analysis.} In the clean system, the spin-polarized Hall conductance $\sigma_{xy}^+$ loses quantization when the conduction and valence bands overlap, because $\sigma_{xy}^+$ of overlapped states are canceled.
%In contrast,  since extended states  are driven to the center of conduction or valence band by disorder, $\sigma_{xy}^+$  is contributed from extended states at the band center in the presence of disorder.
%As a result, spin-polarized Hall conductance $\sigma_{xy}^+$  and thus the spin Chern number restore quantization, giving rise to an emergent $Z_2$ topological invariant carried by the states at the center of the bulk band.
%Due to the bulk-edge correspondence, a pair of helical edge modes becomes robust in the presence of disorder, resulting in a quantized conductance plateau in Fig.\ref{fig1}(b).

\section{Model Hamiltonian}
We consider a $4\times4$ modified disordered BHZ Hamiltonian with the Rashba spin-orbit coupling (SOC): \cite{Andrei2006}
\begin{eqnarray}
 %\nonumber to remove numbering (before each equation)
 H({\bf k}) \!&=&\! \left(
       \begin{array}{cc}
         h({\bf k}) &  \\
              & h^{*}(-{\bf k}) \\
      \end{array}
     \right) + \! H_{R} \! \label{Eq1}\\
 h({\bf k}) &=& (D_x k_x^2 + D_y k_y^2 ) + (m_0 - B_x k_x^2 - B_y k_y^2 ) \tau_z \nonumber\\
 &&+ \hbar (v_x k_x\tau_x  + v_y k_y \tau_y)  + V({\bf r}) \nonumber
\end{eqnarray}
where $h({\bf k})$ and its time-reversal counterpart $h^{*}(-{\bf k})$ act on the spin up and spin down blocks, respectively, and  ${\bf k}$ is the wave vector.  Here $\tau_{x,y,z}$ are the Pauli matrices in the orbital space.
$B_{x(y)}$ and $D_{x(y)}$ describe symmetric and asymmetric parts of the effective masses of the conduction and valence bands in the $x$ ($y$) direction, the mass $m_0$ determines the gap, and $v_{x(y)}$ is the Fermi velocity in the $x$ ($y$) direction.
The long-ranged disorder potential is given by $V({\bf r})=\sum_{n=1}^{N_I}{U_n\exp{[-|{\bf r}-{\bf r}_n|^2/(2\xi^2)}]}$, where $U_n$ is uniformly distributed in $[-W/2,W/2]$ with the disorder strength $W$, and $N_I$ impurities are randomly located at \{${\bf r}_n$\} among $N$ sites \cite{Yanyang2009,Chen2015}. We fix the impurity density $n = N_I /N = 0.2$ and the disorder range $\xi$ = $1$. We note that different $n$ and $\xi$ don't influence our results qualitatively.
We adopt the Rashba SOC term as $H_{R}= V_{R}(k_x s_y - k_y s_x)$, where $V_{R}$ is the strength and $s_{x,y}$ are the Pauli matrices in the spin space.

\begin{figure}[bht]
\centering
\includegraphics[width=3.4in]{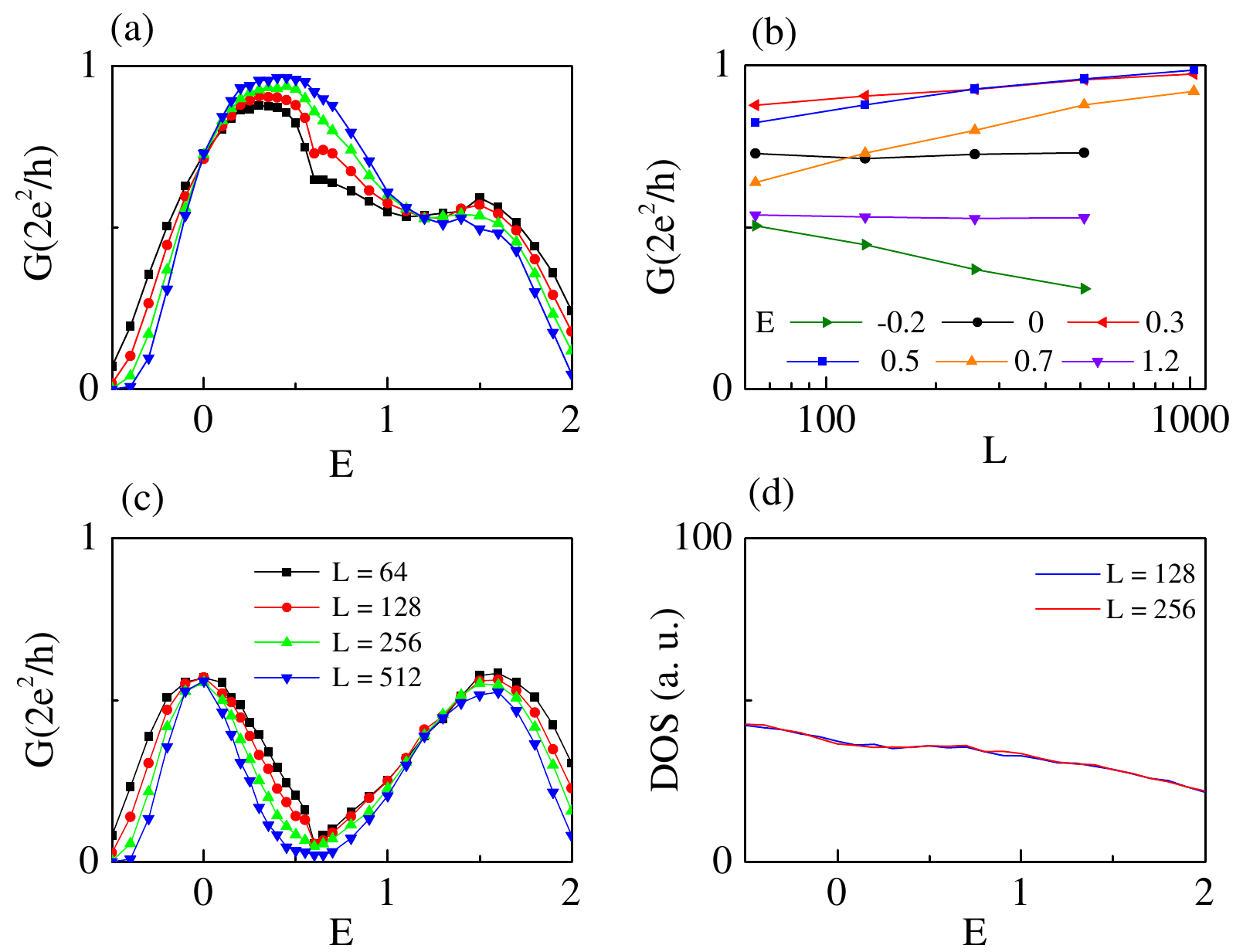}
\caption{(Color online). Two-terminal conductance $G$ versus the Fermi energy $E$ for (a) open (strip sample) and (c) periodic (cylinder sample) boundary conditions in the $y$ direction, respectively. $G$ is averaged over $1000$ disorder configurations with the sample size $L\times L$. (b) Plots $G$ against the sample size $L$ at various Fermi energies $E$ for a strip sample. For comparison, (d) shows DOS in an arbitrary unit (a.u.) as a function of $E$.
The disorder strength is fixed at $W=3.5$.
\label{fig2} }
\end{figure}
If we first ignore the Rashba SOC term $H_R$, then the Hamiltonian $H$ becomes block-diagonal. When $D_{x,y}<B_{x,y}$, the system is a $Z_2 = 1$ QSH insulator with a pair of topological nontrivial helical edge modes for $m_0\cdot B_{x,y}>0$, and it becomes a $Z_2 = 0$ normal insulator for $m_0\cdot B_{x,y}<0$ \cite{Andrei2006}. However, if we tilt the Hamiltonian such that $D_{x}>B_{x}$, the system becomes a topological metal where the helical edge modes are embedded inside the gapless bulk states.
 We note that, unlike the previous three-dimensional topological metals such as Weyl semimetals \cite{Wan2011,Yang2011} and nodal line semimetals \cite{Heikkil2011,Burkov2011}, the  two-dimensional topological metals discussed in the work are not characterized by the topological number in the clean limit, while they are  all featured by the coexistence of bulk and edge states.
 In Fig.\ref{fig1}(a), we show the energy band structure of a nanoribbon of the topological metal with $D_{x}=1.02 B_{x}$. It is clear that the conduction and valence bands overlap so that all the edge modes are completely embedded inside the bulk states. In the numerical simulations, we have discretized the Hamiltonian on a square lattice \cite{Chen2015} with lattice constant $a=1$ and set model parameters as $B_x=B_y=0.5$, $v_x=v_y=0.55$, $D_x=0.51$, $D_y=0$ and $m_0=0.55$.

\section{Disorder-induced quantized edge conductance}

To explore the effects of disorder on the topological metal, we study the two-terminal conductance $G$ by the Landauer-B\"{u}ttiker formula \cite{datta1995}.
In Fig.~\ref{fig1}(b), we plot the two-terminal conductance $G$ as a function of the disorder strength $W$ at the Fermi energy $E=0.5$. The system shows  sizable bulk conductance near $W=0$ because of a large number of conducting channels at $E=0.5$ [see the dashed line in Fig.~\ref{fig1}(a)]. Surprisingly, we find a quantized conductance plateau at $G=2e^2/h$ with increasing the disorder strength $W$ for the open boundary condition in the $y$ direction, while $G$ becomes zero for the periodic boundary condition. The plateau remains quantized until all the states are localized in the strong disorder limit. This strongly suggests the existence of the topological nontrivial helical edge modes. Meanwhile, DOS remains a finite value in the plateau region [see solid blue line in Fig.\ref{fig1}(b)], which rules out the gap reopening by disorder \cite{Shen2009,Groth2009,Jiang2009}.

Now let us perform a finite-size scaling of the two-terminal conductance $G$.
In Fig.\ref{fig2}(a), we plot the conductance $G$ versus Fermi energy $E$ at the disorder strength $W=3.5$ for different sample sizes $L \times L$. There are two critical points near $E_c=0$ and $E_c=1.2$ where $dG/dL=0$ [see also Fig.\ref{fig2}(b)-(c)].
It is found that there exist delocalized edge modes between the two critical points, because the
conductance $G$ increases (decrease) with the system size $L$ for the open (periodic) boundary condition [see Figs.\ref{fig2}(a) and (c)]. These edge modes are robust to disorder and result in a quantized conductance plateau at $G=2e^2/h$ for $0<E<1.2$ in the thermodynamic limit. In Fig.~\ref{fig2}(b), we show the conductance $G$ against the size $L$ for various Fermi energies $E$. The conductance $G$ approaches the quantized value $2e^2/h$ with increasing size $L$ for the energies between two critical points. Therefore, we conclude that the quantized conductance plateau region is determined by the critical points in the bulk band. Furthermore, the system has no band gap because the DOS remains finite for all $E$ as shown in Fig.~\ref{fig2}(d).

\begin{figure}[bht]
\centering
\includegraphics[width=3.5in]{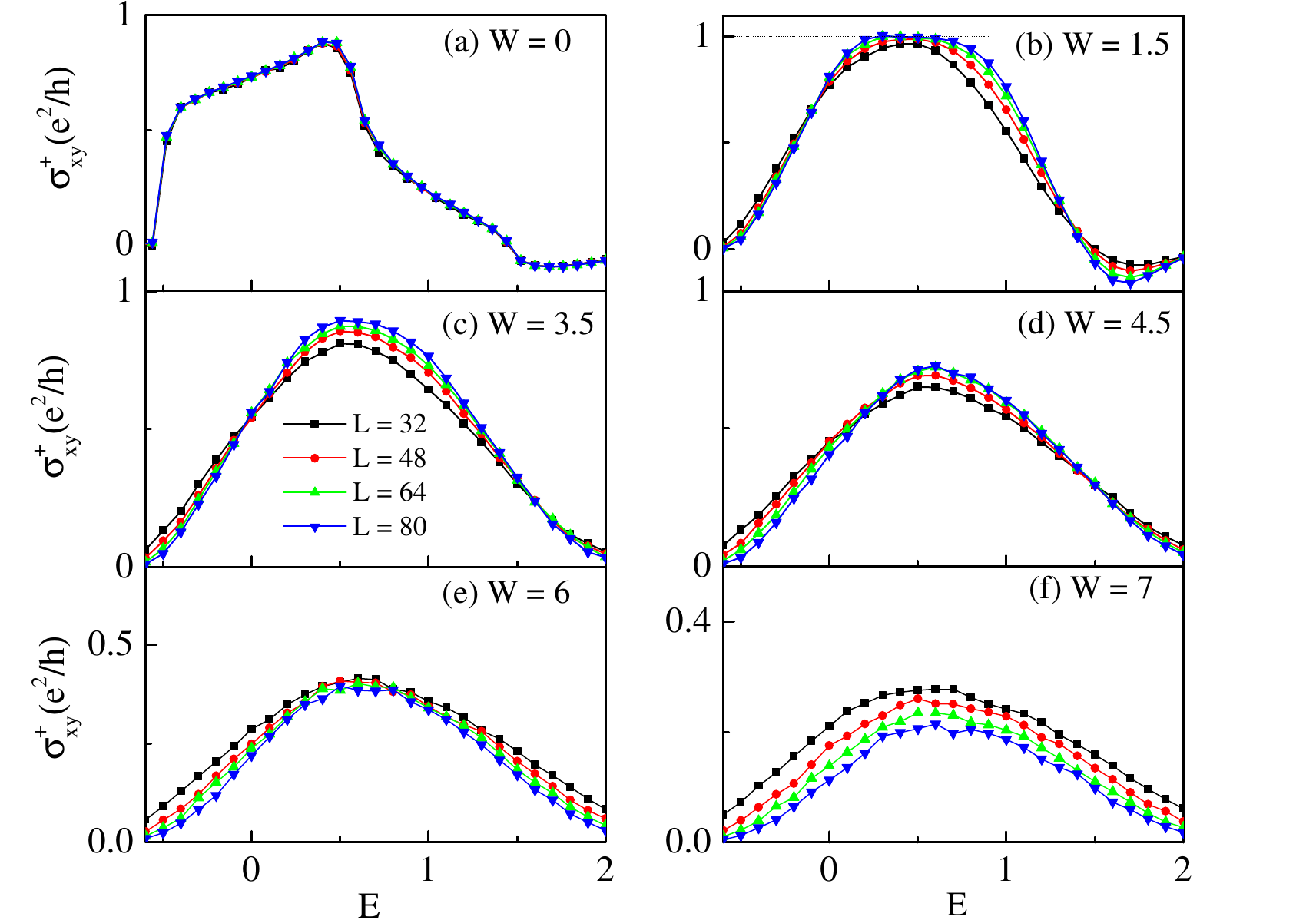}
\caption{(Color online). (a-f) Plots of the Hall conductance $\sigma_{xy}^+$ for the spin-up sector versus the Fermi energy $E$ under various disorder strengths $W$. The colors of lines represent different sample sizes $L\times L$.
\label{fig3} }
\end{figure}

{\em Emergent $Z_2$ topological invariant.}-- Now, we move on to discuss the topology of the bulk states by using the spin Chern number, which will further uncover the topological origin of the quantized edge conductance. Under the time-reversal symmetry, the Hall conductance of spin-down and spin-up sectors obeys $\sigma_{xy}^-=-\sigma_{xy}^+$, and consequently the spin Chern number can be defined by $C_s \equiv (C_{+} - C_{-})/2$~\cite{Sheng2006,Sheng2005,Prodan2009,Prodan2011}. Note that $C_s$ is related to the $Z_2$ classification as $Z_2=C_s$.  We evaluate the Hall conductance $\sigma_{xy}^{\pm}\equiv  C_{\pm}e^2/h $ for the spin-up and spin-down sectors based on a 2D sample of size $L\times L$ under periodic boundary conditions in both $x$ and $y$ directions. The Chern number $C_\pm$ can be evaluated via the non-commutative Kubo formula \cite{Prodan2009,Prodan2011}
% This means that the valence and conduction bands carry a $Z_2$ topological invariant, respectively.
\begin{eqnarray}
  C_{\pm} &=& 2\pi i \langle \Tr[P_{\pm}[-i[\hat{x},P_{\pm}],-i[\hat{y},P_{\pm}]]]\rangle.
\end{eqnarray}
 Here $\langle...\rangle$ represents ensemble-averaged over different disorder configurations, and $(\hat{x},\hat{y})$ denotes the position operators. $P_{\pm}$ is the spectral projector onto the positive/negative eigenvalue of $P s_zP$, while $P$ represents the projector onto the occupied states of $H$. Figure \ref{fig3} shows the scale-dependent behaviors of Hall conductance $\sigma_{xy}^+$  for the spin-up sector versus the Fermi energy $E$ obtained at various disorder strengths $W$.
In the clean limit, the spin-polarized Hall conductance $\sigma_{xy}^+(E)$ is not quantized and independent of the system size for $E$ within the bandwidth [see Fig.~\ref{fig3}(a)]. Therefore, $\sigma_{xy}^+$ is contributed from all the bulk states below the Fermi energy in the thermodynamic limit. On the other hand, in Fig.~\ref{fig3}(b), it is found  that $\sigma_{xy}^+$ scales to a quantized value $e^2/h$ between two scale-independent critical points at $E_c\approx -0.1$ and $1.4$.  Moreover, the slope of the Hall conductance $d\sigma_{xy}^+/dE$ becomes sharper and sharper for larger size $L$ near the two critical points in Fig.~\ref{fig3}(b), in accordance with quantum Hall plateaus transition. For Fig.~\ref{fig3}(b), we expect that $\sigma_{xy}^+$ becomes a step function while $d\sigma_{xy}^+/dE$ is a delta function in the thermodynamic limit, where only the two critical points contribute to $\sigma_{xy}^+$. Therefore, the $Z_2$ topological invariant of the conduction or valence band is carried by the extended states at the critical points. The critical states in the valence and conduction bands are very robust, while they move to the band center and annihilate pairwise in the strong disorder limit [see Figs.~\ref{fig3}(c)-(f)].

\begin{figure}[bht]
\centering
\includegraphics[width=3.2in]{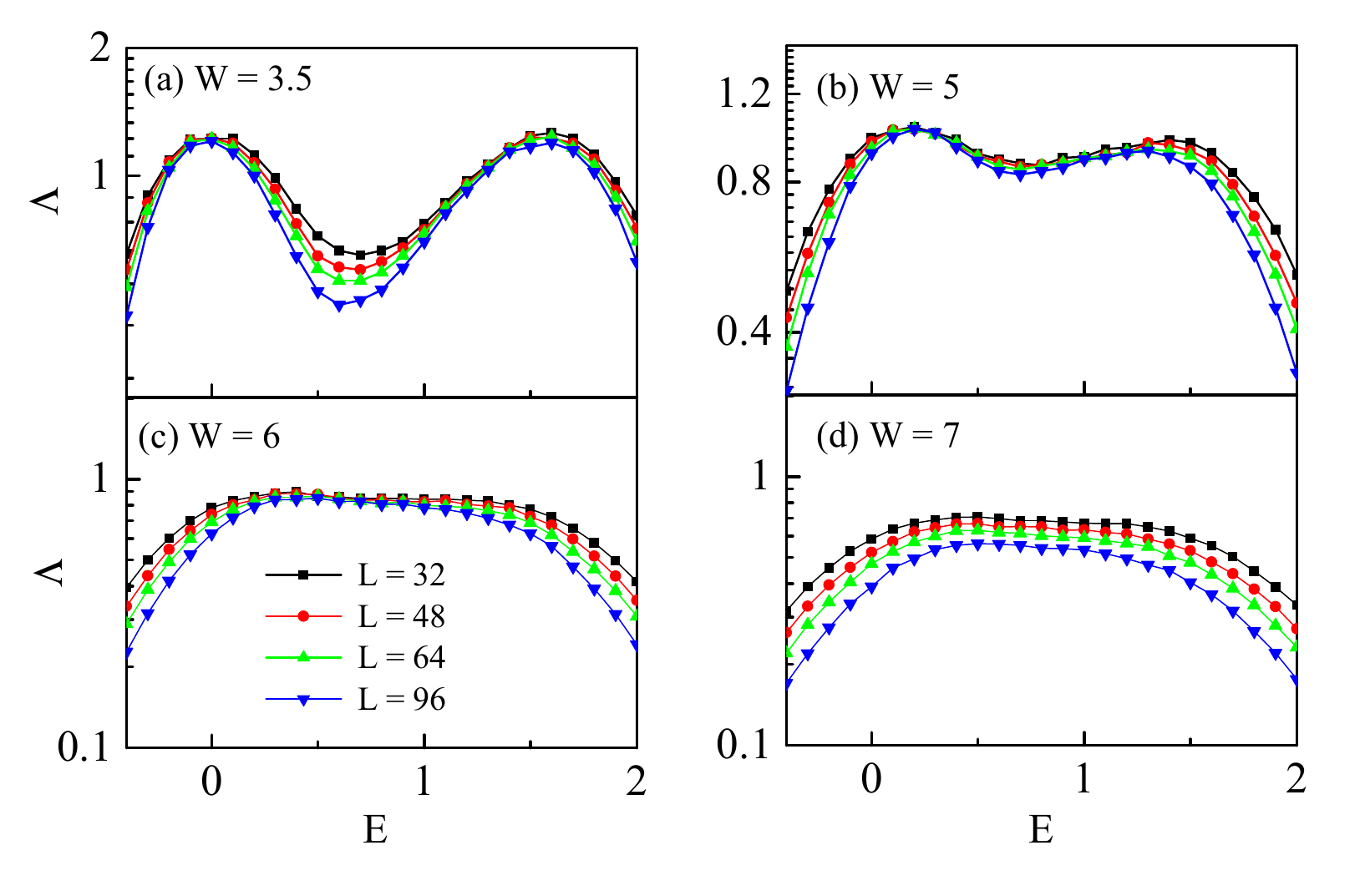}
\caption{(Color online). (a)-(d) The scale-dependent of renormalized localization length $\Lambda$ against $E$ for different $W$.
\label{fig4} }
\end{figure}
Next, we calculate the localization length by considering a 2D cylinder sample  with a periodic boundary condition along the $y$ direction. The length of the cylinder is $L_x$ and the circumference is $L_y=L$.
The localization length $\lambda$ is obtained by the transfer matrix method \cite{MacKinnon1981,MacKinnon1983,Kramer1993}. In general, the renormalized localization length $\Lambda\equiv\lambda/L$ increases with the sample width $L$ ($d\Lambda/dL>0$) in a metallic phase, $d\Lambda/dL<0$ in an insulating phase, and $d\Lambda/dL=0$ at the critical point of  phase transition.
Fig.~\ref{fig4}(a) shows the renormalized localization length $\Lambda$ versus Fermi energy $E$ at the disorder strength $W=3.5$. We find two critical points with $d\Lambda/dL=0$ at $E_c\approx0$ and $1.2$ in the conduction and valence bands, respectively, while all other states are localized with $d\Lambda/dL<0$. With increasing $W$, the two critical points annihilate pairwise in the band center at $W=6$ [see Fig.~\ref{fig4}(c)] and all the states become localized at last [see Fig.~\ref{fig4}(d)]. We emphasize that the consistency between all the phase behaviors determined from the scaling of the localization length, the Hall conductance (in Fig.~\ref{fig3}) and the two-terminal conductance (in Fig.~\ref{fig2}), demonstrates the reliability of the obtained results.

Below we offer an explanation for the main results from a phenomenological point of view. In general, the spin-polarized Hall conductance $\sigma_{xy}^{\pm}$ is contributed from all the bulk states below the Fermi energy in the clean system. It loses quantization when the conduction and valence bands overlap, because the spin-polarized Hall conductance $\sigma_{xy}^{\pm}$  of the overlapped states is canceled. On the contrary,  $\sigma_{xy}^{\pm}$ is contributed from the extended states near the band center in the presence of disorder, because  the extended states are pushed to the center  of the valence or conduction band by disorder.
As a result, $\sigma_{xy}^{\pm}$ restores quantization, giving rise to an emergent $Z_2$ topological invariant carried by the states at the center of the conduction or valence band. Due to the bulk-edge correspondence, there exists a pair of helical edge modes connecting two mobility edges in the valence and conduction bands, which results in the quantized conductance plateau in Fig.~\ref{fig2}(a). Therefore, the helical edge states in the topological metal generally become emergent robust in the presence of disorder.
We note that the topological Anderson insulator (TAI) phase discovered previously originates from band renormalization induced by disorder, which is highly dependent on the disorder types \cite{Shen2009,Groth2009,Jiang2009}. For example, the diagonal Anderson disorder can give rise to TAI, while the spatially correlated disorder \cite{GA2013} and the off-diagonal disorder \cite{Juntao2012} do not. On the contrary, the Z$_2$ topological insulator with a mobility gap discovered in this work results from Anderson localization and thus is independent of disorder types.
\begin{figure}[bht]
\centering
\includegraphics[width=3.4in]{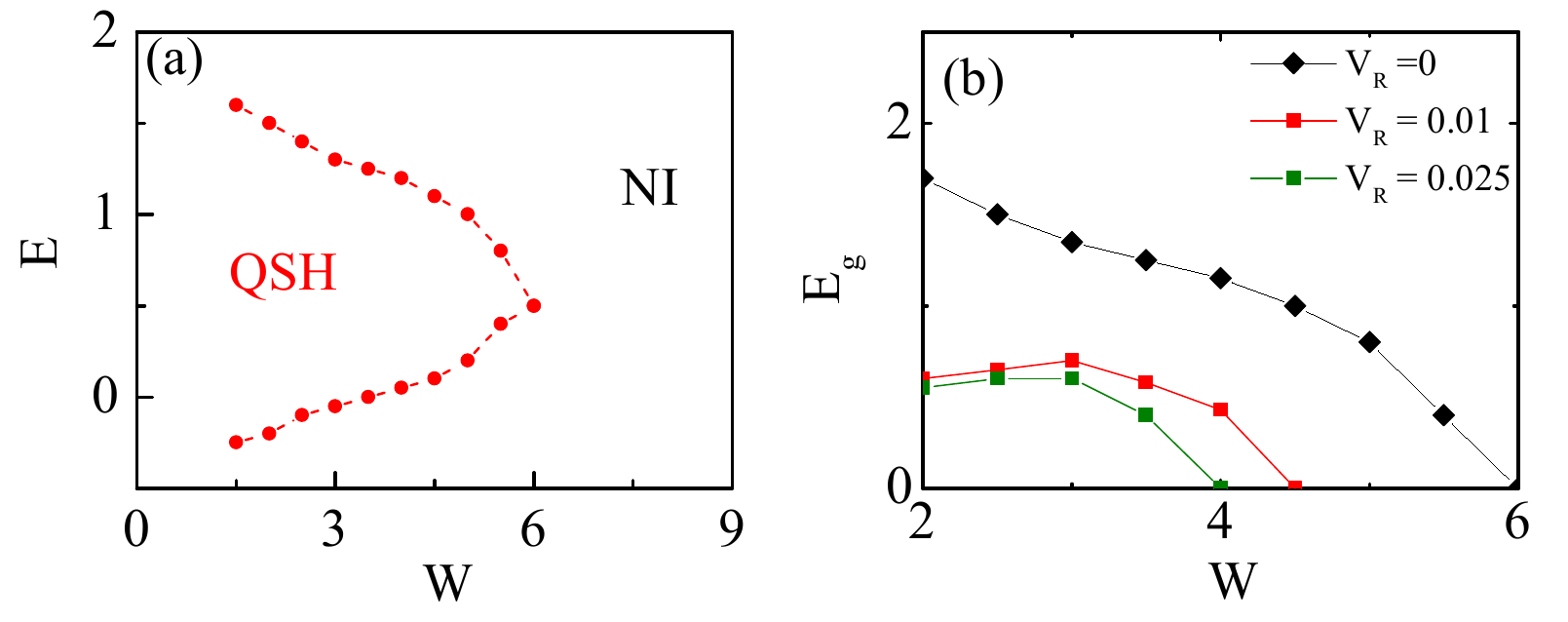}
\caption{(Color online). (a) The phase diagram on the plane of the disorder strength $W$ and Fermi energy $E$ without Rashba SOC ($V_R=0$). The dots are obtained from the localization length scaling and the dashed line is guide to the eye.
(b) The mobility gap $E_g$ as a function of $W$ for various Rashba SOC strengths $V_R$.
\label{fig5} }
\end{figure}

\section{Phase diagram}

 In Fig.~\ref{fig4}(a), we summarize the main results in the phase diagram on the plane of the disorder strength $W$ and the Fermi energy $E$. Although the system is a topological metal in the clean limit at $W=0$, disorder  creates a mobility gap for arbitrary small disorder  by Anderson localization in the thermodynamic limit and drives the system into a QSH insulator. In the absence of Rashba SOC $V_R=0$, the emergent QSH phase can be regarded as two copies of quantum anomalous Hall insulator of unitary class \cite{Altland1997,Schnyder2008}. As shown in Fig.~\ref{fig5}(a), we can see a phase boundary between the QSH insulator and the normal insulator. When the two subblocks of Hamiltonian are coupled by a Rashba SOC term, the system belongs to the symplectic class~\cite{Altland1997,Schnyder2008}. Nevertheless, the mobility gap can still exist. In Fig.~\ref{fig5}(b), we show the mobility gap $E_g$ as a function of the disorder strength $W$ with and without the Rashba SOC.
In the presence of the Rashba SOC, the mobility gap decreases. This is in accordance with the fact that the topologically nontrivial gap of QSH phase is generally reduced by the Rashba term in the modified BHZ model \cite{Yamakage2013}.
We note that the results obtained above are generally applicable to any topological metal systems, including InAs/GaSb QWs and 1T$^{\prime}$-WTe$_2$ monolayer \cite{Liu2008,Qian2014}.  Therefore, we conclude that the topological metals tend to gain a sizable topologically nontrivial mobility gap in the presence of arbitrary small disorder, resulting in the quantized edge conductance of $2e^2/h$ in the thermodynamic limit.

\section{Discussions and conclusion}

 Our results are instructive for varieties of transport experiments in QSH materials such as InAs/GaSb QWs and 1T$^\prime$-WTe$_2$ monolayer~\cite{Du2011,Du2015,Qian2014,Fei2017,Tang2017,Wu2018,Song2018}. It is found that the helical edge transport always coexists with sizable bulk conduction in intrinsic InAs/GaSb QWs \cite{Du2011,Du2014}. On the other hand, when InAs/GaSb QWs are doped by non-magnetic Si impurities, wide conductance plateaus quantized to $2e^2/h$ are observed. The system opens up a mobility gap about $26$K by Si doping  \cite{Du2015}. This strongly suggests the observation of the disorder-induced quantized edge transport phenomena by creating a mobility gap in the bulk, just as we stated above.
More recently, STM measurements demonstrate that 1T$^\prime$-WTe$_2$ monolayer has a metallic bulk band structure as initially predicted by {\it ab inito} calculation \cite{Qian2014,Song2018}. However, the 1T$^\prime$-WTe$_2$ is confirmed to support quantized edge conductance with insulating bulks  \cite{Fei2017,Wu2018}. Surprisingly, the QSH effect in this material is reported to survive at temperatures even up to 100 Kelvin, meaning the existence of a large insulating gap, which cannot be explained by the current theories \cite{Qian2014,Fei2017,Tang2017,Song2018,AM2016,Wu2018}.
Our results of disorder-induced large mobility gaps in topological metals provide a new viewpoint to reconcile this controversial issue.

In summary, we have studied the disorder effect in topological metals and found disorder can always lead to a quantized edge transport protected by an emergent $Z_2$ invariant. In the clean system, the spin-polarized Hall conductance $\sigma_{xy}^+$ loses quantization when the the conduction and valence band overlap. On the contrary, $\sigma_{xy}^+$ is contributed from extended states at the center of the conductance or valence band in the presence of disorder.
As a result, the system restores a quantized spin Chern number $C_s=1$ and thus turns into a $Z_2=1$ QSH insulator with robust helical edge modes transport. Our work can explain recent experiments of realizing QSH effect in disordered topological metals.

\section{Acknowledgement}
We thank Haiwen Liu and Rui-Rui Du for illuminating discussions.  This work is financially supported by NBRPC (Grants No. 2015CB921102), NSFC (Grants No. 11534001, No. 11822407, No. 11704106, No. 11974256), and supported
by the Fundamental Research Funds for the Central Universities.
H. Jiang and C.-Z. Chen are also funded by the Priority Academic Program Development of Jiangsu Higher Education Institutions and NSFC of Jiangsu province BK20190813.
D.-H.X. is also supported by the Chutian Scholars Program in Hubei Province.

%\bibliographystyle{apsrev4-1} % Tell bibtex which bibliography style to use
%\bibliography{Bulk_edge}

\end{document}